\documentclass[11pt]{article}
\usepackage{paspconf,epsf}

\def\msun{{\rm\,M_\odot}}
\def\msun{{\rm\,M_\odot}}
\def\kpc{{\rm\,kpc}}
\def\spose#1{\hbox to 0pt{#1\hss}}
\def\lta{\mathrel{\spose{\lower 3pt\hbox{$\mathchar"218$}}
     \raise 2.0pt\hbox{$\mathchar"13C$}}}
\def\gta{\mathrel{\spose{\lower 3pt\hbox{$\mathchar"218$}}
     \raise 2.0pt\hbox{$\mathchar"13E$}}}
\let\simless=\lta

\def\sech{\mathop{\rm sech}\nolimits}

\begin{document}

\title{Effect of the Magellanic Clouds on the Milky Way disk\\
  and VICE VERSA}

\author{Martin D. Weinberg}
\affil{Department of Physics \& Astronomy, University of Massachusetts, Amherst, MA 01003-4525, USA}

\begin{abstract}
  The satellite-disk interaction provides limits on halo properties in
  two ways: (1) physical arguments motivate the excitation of
  observable Galactic disk structure in the presence of a massive
  halo, although precise limits on halo parameters are
  scenario-dependent; (2) conversely, the Milky Way as a whole has
  significant dynamical effect on LMC structure and this interaction
  also leads to halo limits.  Together, these scenarios give strong
  corroboration of our current gravitational mass estimates and
  suggests a rapidly evolving LMC.
\end{abstract}

\section{Introduction}

Previous attempts at disturbing the Galactic disk by the Magellanic
Clouds relied on direct tidal forcing.  However, by allowing the halo
to actively respond rather than remain a rigid contributor to the
rotation curve, the Clouds may produce a wake in the halo which then
distorts the disk.  I will describe this dynamical interaction and
present results based on both linear theory and n-body simulations
(\S\ref{sec:WAKE}).
  
Even without a massive satellite, these same physical processes may be
responsible for exciting disk structure.  For example, persistent halo
structure may result from discrete blobs, dark clusters and poorly
mixed streams of material within the halo or from past fly-bys or
minor merger events.
  
Finally, the tables turned, the Milky Way halo has a profound effect
on LMC structure (\S\ref{sec:LMC}).  Dynamical arguments lead to a
consistent determination of LMC and Milky Way Halo mass and a
prediction of the LMC stellar halo by tidal heating.  A fractionally
more massive spheroid has implications for microlensing optical depth.

\section{Scenario \#1: LMC disturbs Halo $\rightarrow$ Halo disturbs Disk}
\label{sec:WAKE}

\subsection{Mechanism overview}

\begin{figure}[ht]
\mbox{
  \mbox{\epsfxsize=2.6in\epsfbox{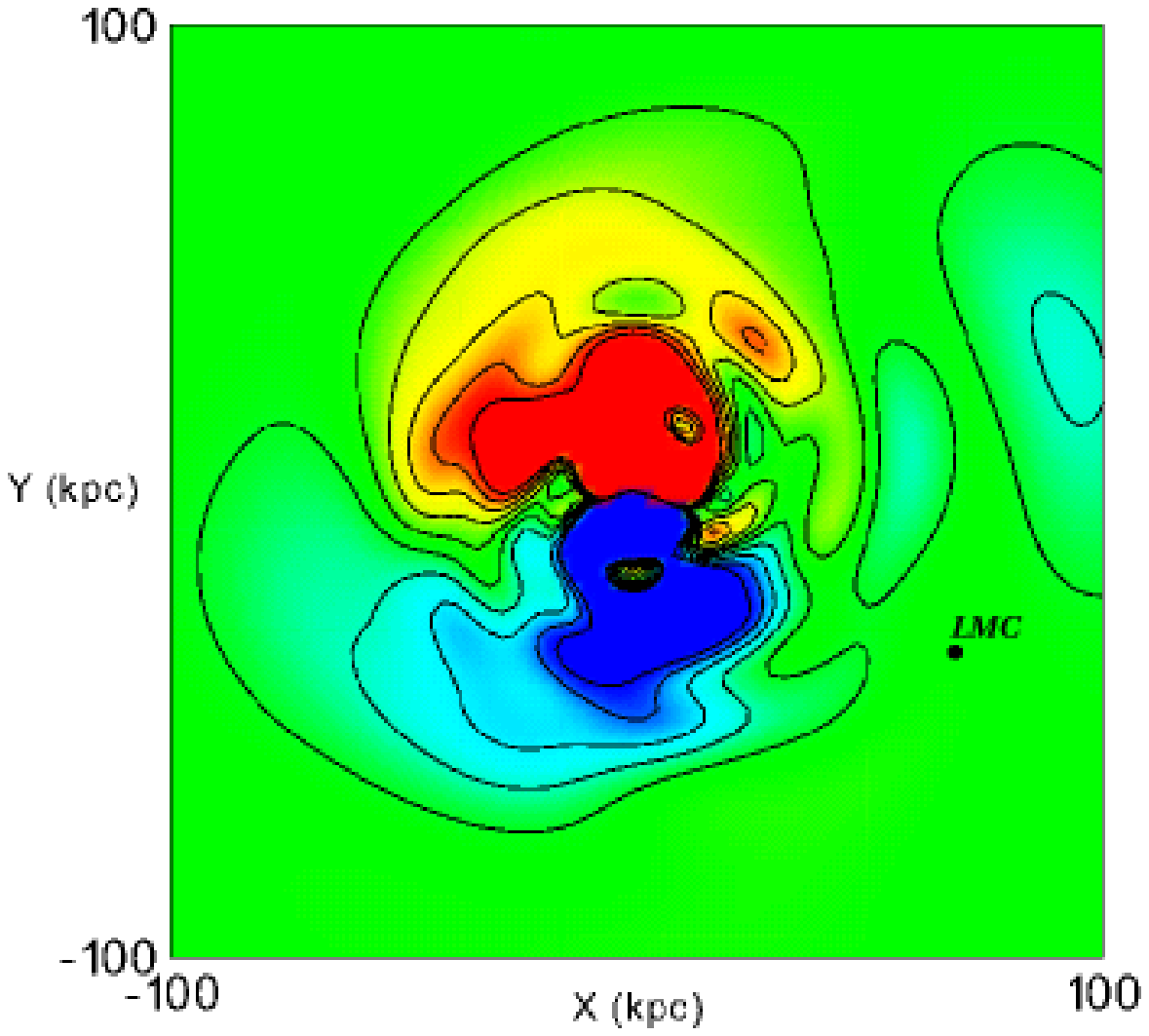}}
  \mbox{\epsfxsize=2.6in\epsfbox{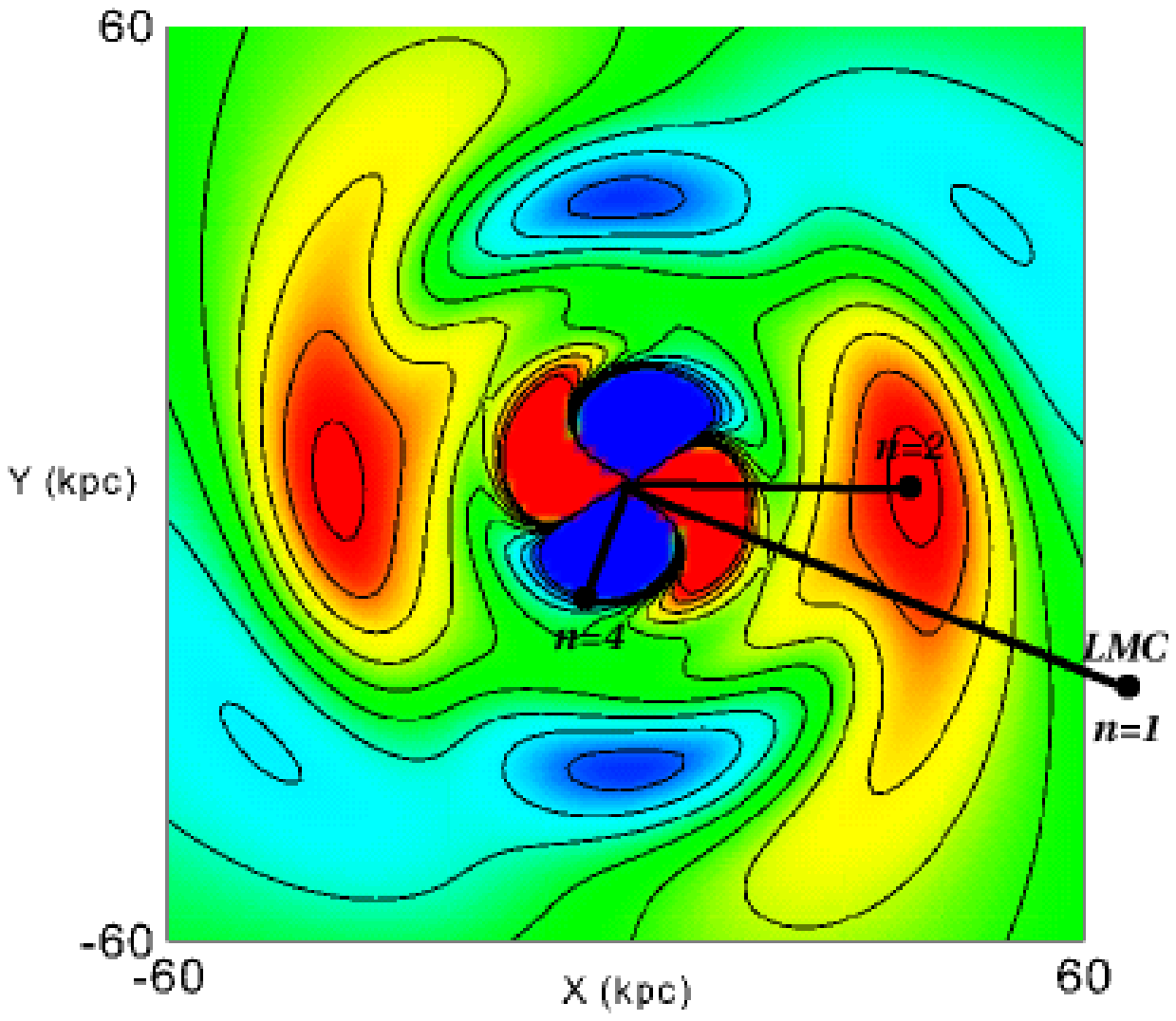}}
}
\caption{Wake from n-body simulation for Milky Way and LMC
  interaction.  Left: density wake due to LMC-like satellite viewed in
  orbital plane.  The negative (positive) overdensity is color deep
  red (deep blue). The peak relative density is $\approx15\%$. Current
  orbital radius is nearly that of a circular orbit with the same
  energy.  Right: Blow up showing quadrupole component only.  The peak
  relative density is a factor of two smaller than in the left-hand
  panel.  The radial location of the $n=1,2,4$ resonances are shown.
  These features agree nicely with the analytic predictions (Weinberg
  1998).}
\label{fig:halo}
\end{figure}

The basic physics is well-understood: a perturbation external to a
galaxy excites a wake in the halo.  For a satellite, this perturbation
is outside the disk but peaks in the halo near the satellite location.
The gravitational force of wake then attracts the satellite; this is
dynamical friction.  An n-body example of this wake is shown in Figure
\ref{fig:halo}.

Although the peak in {\em relative} density is near the perturber, the
quasi-periodic satellite orbit resonates with orbits at
higher-frequency and smaller radii.  This permits harmonics of the
primary response to produce a non-local wake in the inner halo.  For
example, let us consider the LMC orbit.  The characteristic orbital
frequency is $\Omega_p \equiv V_o/ r_{mean}$; in other words, choose
$r_{mean}$ to yield the correct azimuthal orbital frequency for the
flat rotation curve with value $V_o$.  The $n:1$ resonances are
defined by $m\Omega - \Omega(r)=0$ or at characteristic radii given by
$r\sim r_{mean}/n$.  For $r_{mean}\approx70\kpc$ (cf. LMC orbit), the
$n=2,4$ harmonics are at 35 and 17.5 kpc, respectively (see Fig.
\ref{fig:halo}).  The lowest radial and angular orders have the
largest amplitudes.  In particular, the $m\simless2$ angular harmonics
have the range and amplitude to measurably distort the disk as will be
described below.

\subsection{Some subtleties: modes and resonances}

As in any dynamical system, the evolution of a galaxian halo to a
perturbation depends on its underlying modal structure.  In stellar
systems, there are both discrete and continuous modes.  The continuous
modes are excited as a sort of wave packet which disperses though
phase mixing.  In some cases, the discrete modes will dominate the
response.  For a dynamically stable halo, these are {\em damped}
modes.  Because all responses are superpositions of the same modes,
the response will be similar in overall appearance for widely varying
perturbations.

The pattern speeds of these lowest-order modes tend to be very small.
There is a good reason for this: this frequency must be in a range
relatively devoid of commensurabilities with the internal stellar
orbits.  As a consequence they respond most strongly to low-frequency
external forcing.  Relative to characteristic inner galaxy orbital
frequencies, the frequencies of a distant orbiting satellite are quite
small indeed, and therefore can easily drive the low-order natural
modes.  In this case, we expect the modes to be entrained by and
follow the position angle of the disturbance.

\subsection{Consequences}

In the case of the LMC--Galaxy interaction, the halo wake and disk
modes conspire to produce structure as follows:

\begin{figure}[ht]
  \mbox{\epsfxsize=3.5in\epsfbox{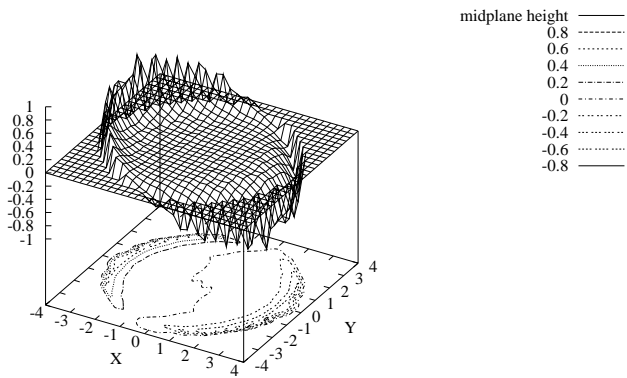}}
  \caption{Warp in Milky Way disk due to LMC.  Each unit in $X$ and
    $Y$ is 7 kpc.  Each unit in height is 2.1 kpc.  From Weinberg
    (1998).}
  \label{fig:warp}
\end{figure}

\begin{enumerate}
\item The $l=2$ halo distortion due to a satellite on a polar
  orbit---such as the LMC---can excite an $m=1$ vertical disturbance
  in the disk: {\em a warp}.  The relative density in the
  satellite-induced halo distortion near the outer disk is roughly
  $5\%$.  Although this is only enough for a low-amplitude warp, the
  disk response, is also a superposition of modes and the disk warping
  is dominated by the excitation of discrete bending modes.  If the
  wake pattern frequency is approximately commensurate with the disk
  bending frequency, a significant amplification can result in
  kiloparsec-scale warps.
  
  As an aside, for n-body simulations with fewer than several hundred
  thousand halo particles, the Poisson noise is sufficient to raise
  the low-order modes to the same amplitude.  The difference here is
  that the phase is stochastically varying in time whereas the
  periodic forcing yields the resonant amplification.
  
  Figure \ref{fig:warp} shows an example for Milky Way--LMC
  parameters.  The warp follows the halo distortion and therefore has
  a {\em retrograde} pattern speed.

\begin{figure}[ht]
\mbox{
  \mbox{\epsfxsize=2.6in%
    \epsfbox{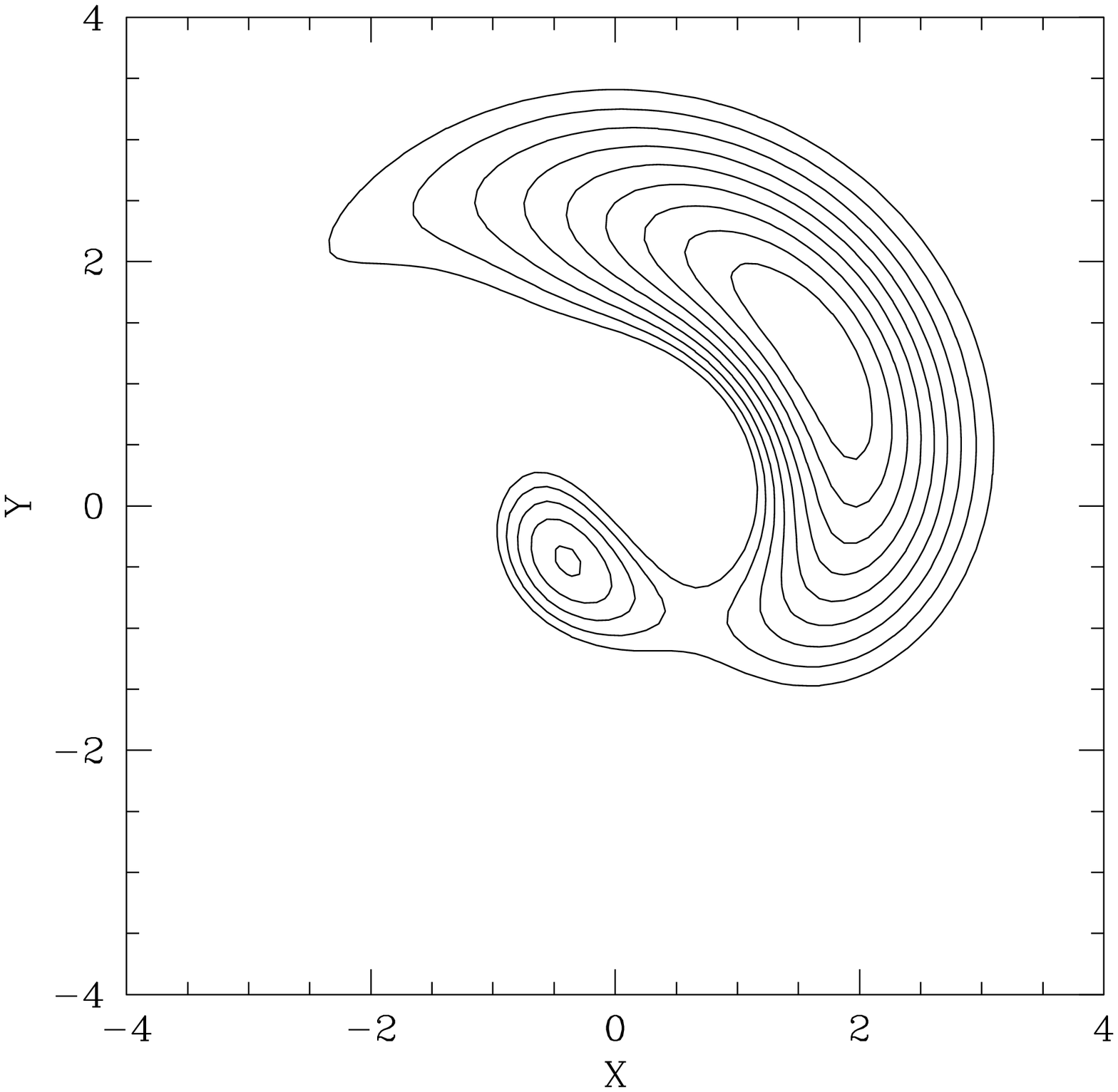}}
  \mbox{\epsfxsize=2.6in%
    \epsfbox{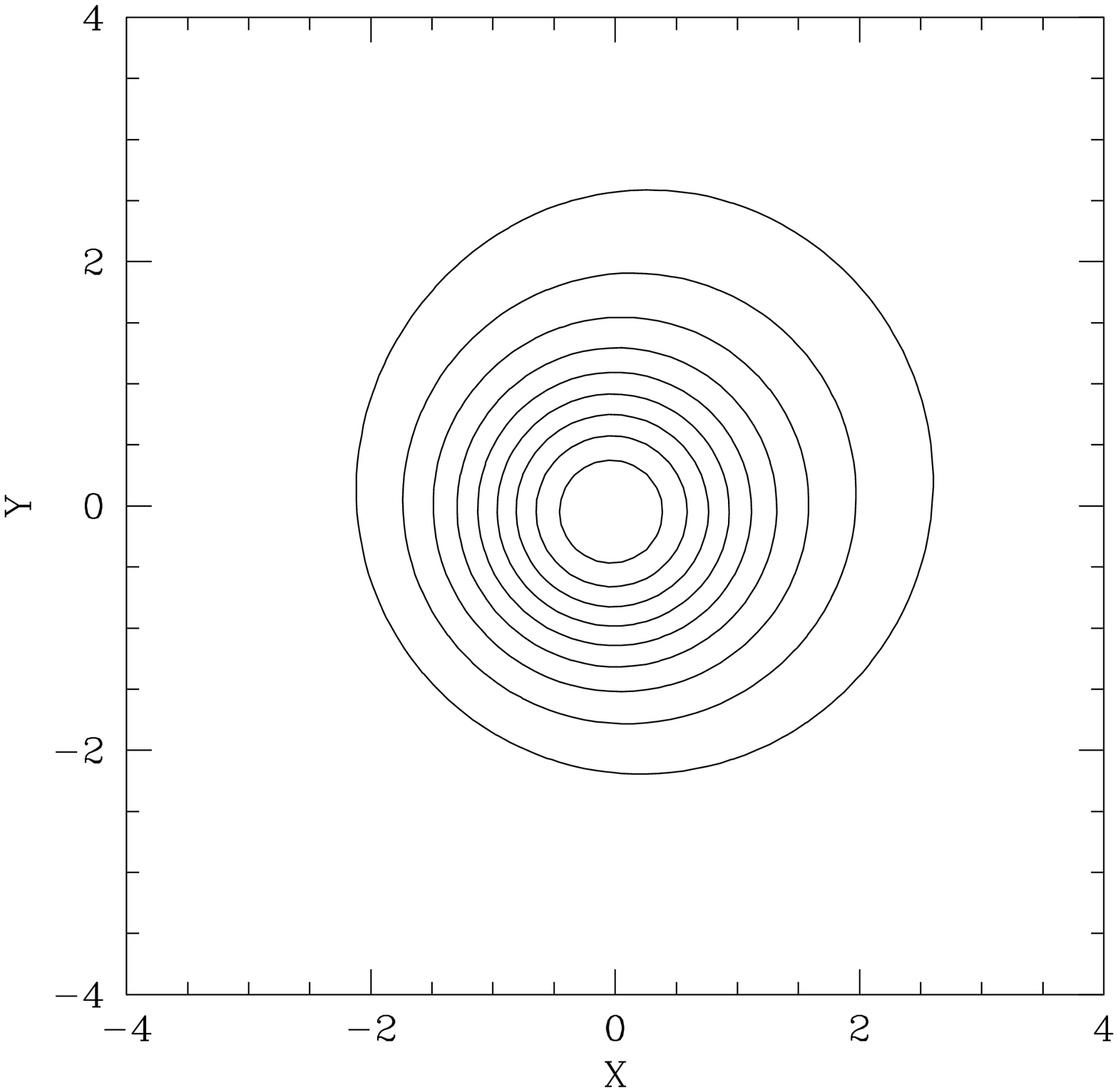}}
  }
\caption{Predicted dipole distortion in Milky Way caused by LMC halo
  wake.  Left: $m=1$ in-plane distortion Right: isodensity profiles
  for large LMC mass ($M=2\times10^{10}\msun$, see
  \S\protect{\ref{sec:LMC}}).}
\label{fig:radvel}
\end{figure}

\item The $l=1$ in-plane distortion can be quite large although it
  produces no warp\footnote{It is straightforward to convince yourself
    of this using the form of the multipole expansion and symmetry
    considerations.}.  Both theory and N-body simulations predict
  dish-like vertical distortions resulting from $l=1$ disturbances.
  Also significant, the $m=1$ distortion in the Galactic plane is at
  the 10--20\% level although the corresponding change in radial
  velocity is rather small (Fig. \ref{fig:radvel}).
  
\end{enumerate}
A vertical warping presents a {\em vertical} velocity distortion as a
diagnostic.  Earlier this year Smart et al. (Smart et al. 1998)
inferred the nearby vertical velocity distortion using proper motions
of O-B stars.  The prediction from the model has a similar trend to
the Smart et al.  inference but lower amplitude.

\subsection{Some n-body surprises}

\begin{figure}[ht]
\vbox{
  \hbox{\epsfxsize=5.0in\epsfbox{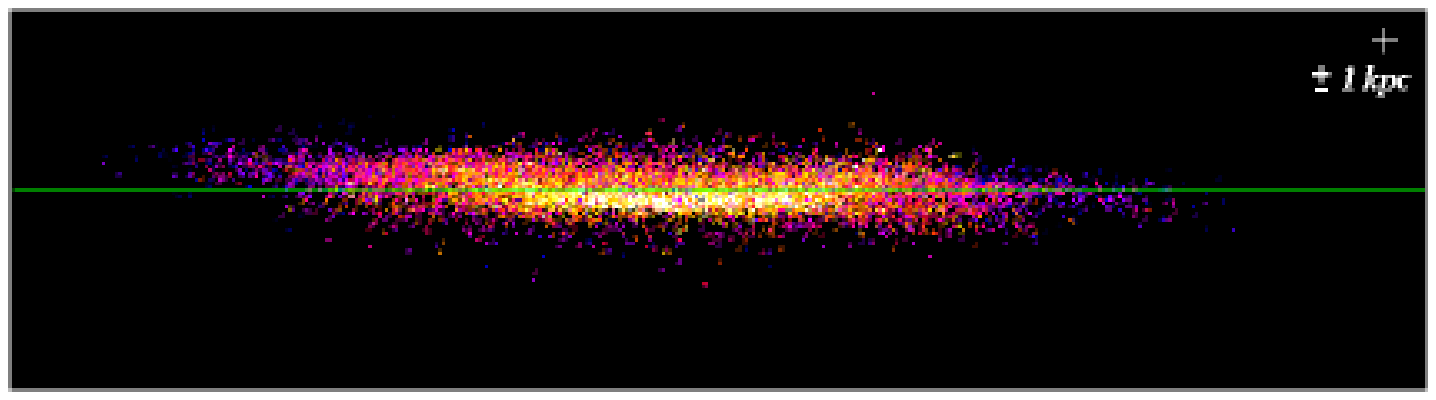}}
  \hbox{\epsfxsize=5.0in\epsfbox{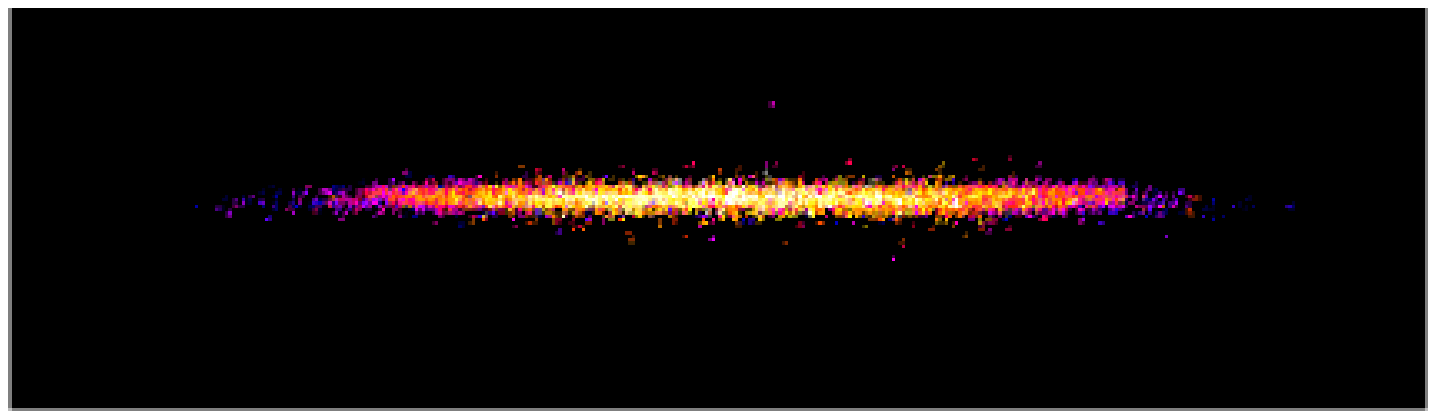}}
}
\caption{N-body simulation of LMC-Milky Way interaction (N=500,000).
  The satellite is approaching pericenter below the plane. Upper: edge
  on view of warped disk.  The inset cross shows $\pm1\kpc$.  Lower:
  edge of view of the using same initial conditions but without
  satellite.
  }
\label{fig:nbwarp}
\end{figure}

Figure \ref{fig:nbwarp} shows the results of n-body simulation of a
disk-halo with and without a satellite perturber.  A recent paper by
Velasquez \& White (1998) describes a warp excitation using rings.
Although I have not made a detailed comparison, their warp appears
consistent with these predictions.

My previous analysis has concentrated on the $l=2$ distortion because
of its role in producing the ``integral sign'' warp.  However, as
Figure \ref{fig:halo} clearly shows, the $l=1, m=1$ component is
significant.  The first panel in the figure shows the $l\ge1$
harmonics is dominated by the inner $m=1$ component.  In fact, this is
due to the existence of a weakly damped $m=1$ halo mode (Weinberg
1994).  Since the LMC orbit is roughly polar, the inner galaxy is
periodically accelerated northward and southward on a roughly gigayear
time scale.  This leads to both a corrugated bending and disk heating
(see also Edelsohn \& Elmegreen 1997).  The disk heating in Figure
\ref{fig:nbwarp} may be exacerbated by particle noise.

\subsection{Summary: limits on the halo mass and profile}

\begin{enumerate}
\item The halo wake is not the only possible mechanism for producing
  warps (Binney 1992, Binney et al. 1997, Jiang \& Binney 1998).
  Nonetheless, it appears the LMC can have observable effects on the
  Milky Way disk.  The magnitude of these effects depends in part on
  the mass of the LMC and we will explore this in detail below.
  
\item For a Milky Way halo mass of $5\times10^{11}\msun$ inside of 50
  kpc, the LMC can both produce the magnitude of the observed warp and
  arrange observable $m=1$ distortions (e.g. the inner-disk offset and
  a predicted outer disk offset).
  
\item A less massive halo than the 10:1 ratio is inconsistent with the
  rotation curve and leads to disk instability.  Nonetheless, similar
  warp amplitudes persist until about 7.5:1 beyond which the effect of
  the wake on the disk decreases.
  
  However, the halo-disk interaction is diminished if the halo mass is
  increased by 50\%--100\%.  As the halo mass increases, the orbital
  frequency increases.  This detunes the disk bending mode and
  decreases the excitation of the inner $m=1$ mode.
  
\end{enumerate}

\section{Scenario \#2: Halo disturbs LMC}
\label{sec:LMC}

A reliable estimate of the disk effects due to LMC disturbance
requires a reliable estimate of the LMC mass and orbit.  Recent proper
motion measurements (Jones et al. 1994) combined with simulations has
helped secure the orbital parameters.  The standard mass estimates,
however, vary by a factor or three!  In addition, the time-dependent
tidal forcing of the LMC along its orbit has dramatic consequences for
the internal dynamics of the LMC.  I will describe initial
investigations into both below. These avenue of inquiry leads to firm
estimates.

\subsection{Mass \& Structure of LMC}

There have been a wide variety of LMC studies, most of which treat the
LMC as a separate galaxy and use the standard mass and mass density
estimates: rotation curves, star counts, surface brightness profiles.

Two relatively recent and often-cited rotation curve studies,
Meatheringham et al.  (1988) and Schommer et al. (1992), estimate LMC
masses of $6\times10^9\msun$ and $1.5\times10^{10}\msun$.  The main
difference between these two is not the values of $V_c$ but the radial
extent of the curve.

Alternatively, from the Milky Way's point of view, the LMC is an
oversize globular cluster.  Its tidal radius is measurable and depends
on both the Milky Way rotation curve and the LMC mass (and, weakly,
its profile).  This gives us additional checks and limits on both the
Milky Way halo profile and the LMC mass.

\subsubsection{The LMC tidal radius and mass}

\begin{figure}[ht]
  \mbox{\epsfxsize=3.25in\epsfbox{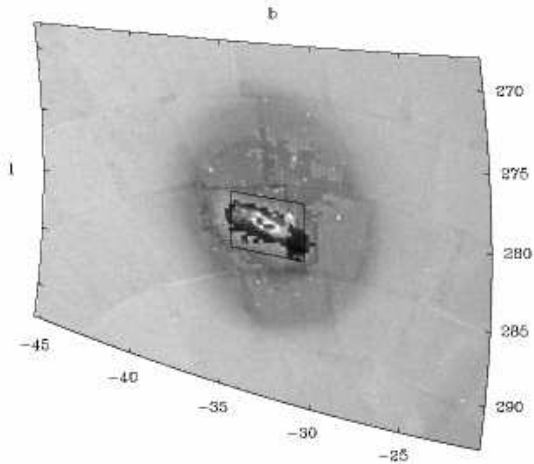}}
  \caption{Stars from the LMC field from USNO-A1.0 catalog, binned on $512\times512$ Cartesian grid
    centered on the LMC. Grey scale is logarithmic from zero (white) to 632 (black).} 
  \label{fig:fig1}
\end{figure}

Figure ~\ref{fig:fig1} shows the binned star counts from the digitized
POSS in the USNO-A catalog (Monet 1996).  The extended stellar
population extends outward to at least $10^\circ$ from the center (see
also Irwin 1991).  This population has M-giant colors.  Its extent
motivates a tentative identification with a halo or spheroid rather
than the standard LMC disk.  I will discuss a plausible dynamical
explanation for this population at the end.  From the extent of the
stellar halo of the LMC, we can estimate the tidal radius and the mass
of the Cloud.  This was attempted by Nikolaev \& Weinberg (1998) based
on the USNO-A1.0 data.  However, large extinction in B-band and poor
quality of photometry ($\sim 0.5^m$) made it difficult to obtain the
structural parameters of the LMC accurately.

The 2MASS survey, which operates in near-infrared where the
interstellar extinction is much smaller and which has much better
photometric accuracy allows more accurate determination of the
structural parameters.  During the period from 03.19.98 to 04.02.98,
the 2MASS southern facility observed five $6^\circ\times1^\circ$
fields of the Large Magellanic Cloud. The fields were selected to
cover both central regions and the periphery ($\sim 5^\circ$) of the
LMC.  The positional accuracy of the scans is $\sim 1$ arc second and
the photometric error is $\sim 0.03^m$.  For this structural analysis
we use only $K_s$-band data; additional colors can boost the
sensitivity to radial density profile.  We select 12 subfields
$0.5^\circ\times0.5^\circ$ in size which probe the LMC halo at the
projected radii of $2^\circ - 5^\circ$ from the LMC center
($l_{II}=280.5^\circ, b_{II}=-32.9^\circ$). The general approach may
be extended to very large datasets by hierarchical partitioning.  Both
this and the multicolor analysis is in progress.

We use Gaussian and power-law spherical models to describe spatial
density of the LMC: $\rho \propto e^{-{r^2 /2 a^2}}$ and $\rho \propto
\left( 1 + r^2/a^2 \right)^{-\gamma}$.  The power-law form was used by
Elson, Fall \& Freeman (1987) to fit the profiles of the globular
clusters in the LMC, who derived $\gamma \approx 1.3$ for their sample
of 10 rich clusters.  We tried both fixed exponent model ($\gamma =
2.0$) and models where $\gamma$ is a free parameter.  We have also
attempted triaxial models but invariably found that there is a
degeneracy between the scale length along the line-of-sight and the
luminosity function.  Our multicolor analysis should also break this
degeneracy.

To estimate the mass of the LMC, we fit the profiles to King models to
estimate the tidal radius.  Standard arguments then suggest
\begin{equation}
M_{LMC} = \left( r_t \over R_{LMC} \right)^3 2 M_{MW},
\end{equation}
where $R_{LMC}$ is the distance to the LMC and $M_{MW} = 5 \times
10^{11}\;M_\odot$ is the mass of the Milky Way.  The mass of the LMC
which follows from this expression includes both the halo and the disk
mass.  This procedure will underestimate the mass for two reasons.
First, simulations suggest that the observed $r_t$ is 75\%--80\% of
the critical point.  Second, a tidally-limited object is likely to be
elongated toward the Galactic center and therefore roughly along the
line of sight.  For a centrally-concentrated object, the axis ratio is
$a/c=1.5$.  Conservatively, the first correction yields a factor of
$(10/8)^3\approx 2$.  The second increases the enclosed volume by
roughly $3/2$ but whether or not this should be included depends on
orientation.  A reasonable correction factor is then between 2 and 3
and we conservatively choose the former.

The parameters of the best-fit models and the corresponding lower
limits on the LMC mass are presented in Table \ref{tab:table1}.  The
bracketed term in the final column is the inferred mass with no tidal
radius or orientation correction.
\begin{table}[ht]
  \caption{Structural Parameters of the LMC Halo}
  \label{tab:table1}
  \begin{center}
    \begin{tabular}{|l|c|c|c|} 
      \hline
      \em Model & \em Scale Length & $r_t$ (kpc) & $M_{LMC}$ \\\hline
      Gaussian @ 50 kpc & $2.64 \pm 0.04 \; kpc$ & $10.8$ & $2.0 [1.0] \times 10^{10} M_\odot$ \\
      Power-Law ($\gamma = 2.0$)  & $2.85 \pm 0.08 \; kpc$ & $10.8$ &
      $2.0 [1.0] \times 10^{10} M_\odot$ \\
      \hline
    \end{tabular}
  \end{center}
\end{table}
Both the profiles and the tidal radii are in excellent agreement with
each other and suggest that the mass of the LMC is about $2.0 \times
10^{10} M_\odot$.  If the disk mass inside $3^\circ$ is about $4
\times 10^9\;M_\odot$ (Meatheringham et al. 1988, De R\'ujula et al.
1995), then the LMC halo mass must be greater than $1 \times
10^{10}\;M_\odot$.

As an independent check, we make a crude estimate of the mass of the
LMC from the analysis of the halo population using the star counts in
our fields.  Most of the sources observed by 2MASS are M-giants with
the absolute magnitude in $K$-band $K < -4^m$ (for the distance to the
LMC of $50\,\kpc$ and 2MASS $K_s$-band flux limit of $14.3^m$).
Assuming that these M giants are representative of an intermediate age
population with a spheroidal distribution, we may estimate the total
stellar mass using an infrared luminosity function.  For this purpose,
we adopt a the Galactic luminosity function in Wainscoat et al.
(1992).  Integrating over the luminosity function with a standard
luminosity-mass relation results in stellar mass of $\approx 4 \times
10^9\;M_\odot$, which is consistent.

\subsubsection{Rotation curves: another consistency check}

Schommer et al. (1992) summarizes the derived rotation curve for
clusters, planetary nebulae and H{\sc I} including the Meatheringham
et al. results (see Fig. 8 from Schommer et al.).  Using a luminosity
function derived for an exponential disk and typical velocity
dispersion in halo with a flat rotation curve, one finds that the
circular velocity $V_c$ is roughly 20--30\% larger than the rotation
value $V_o$. For a rotation curve with $V_o\approx75$ km/s, one finds
a mass within 10.8 kpc of $M\gta1.4\times10^{10}\msun$ which is nicely
consistent with the tidal radius estimate.  See Schommer et al. for
more extensive discussion.

This consistency between the apparent tidal radius of the LMC and the
dynamical mass estimate of Milky Way (which is dominated by the dark
halo at $R_{LMC}$) is comforting for the dynamicist who uses Newton's
Law of Gravity on scales at least $10^8$ times larger than the direct
solar system tests.

\subsection{Milky Way heating of the LMC}

\begin{figure}[ht]
  \mbox{\epsfxsize=4.5in\epsfbox{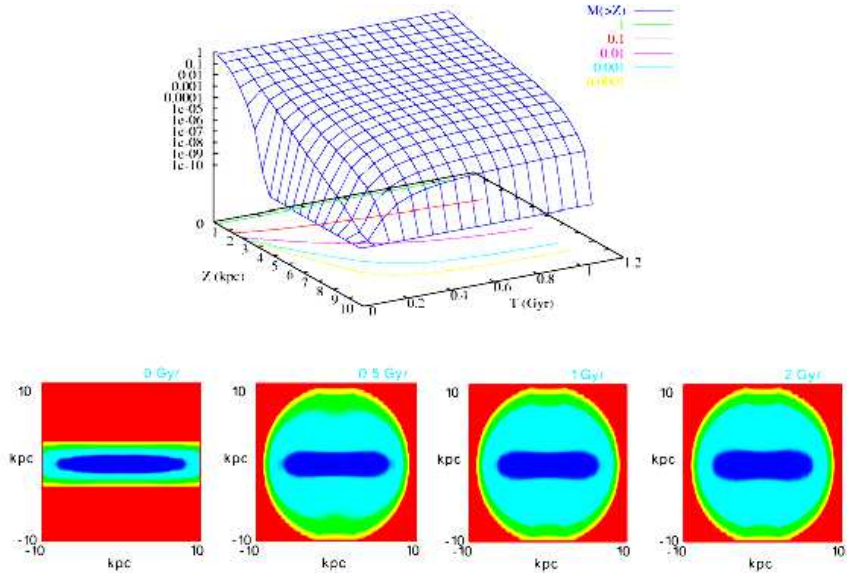}}
  \caption{LMC disk heating by the Milky Way.  The top panel shows the
    cumulative distribution of material at a height $Z$ or larger
    above the disk plane.  The lower four panels show the projected
    surface density distribution for the edge on view of disk.}
  \label{fig:lmcdisk}
\end{figure}

\begin{figure}[th]
  \mbox{\epsfxsize=4.5in\epsfbox{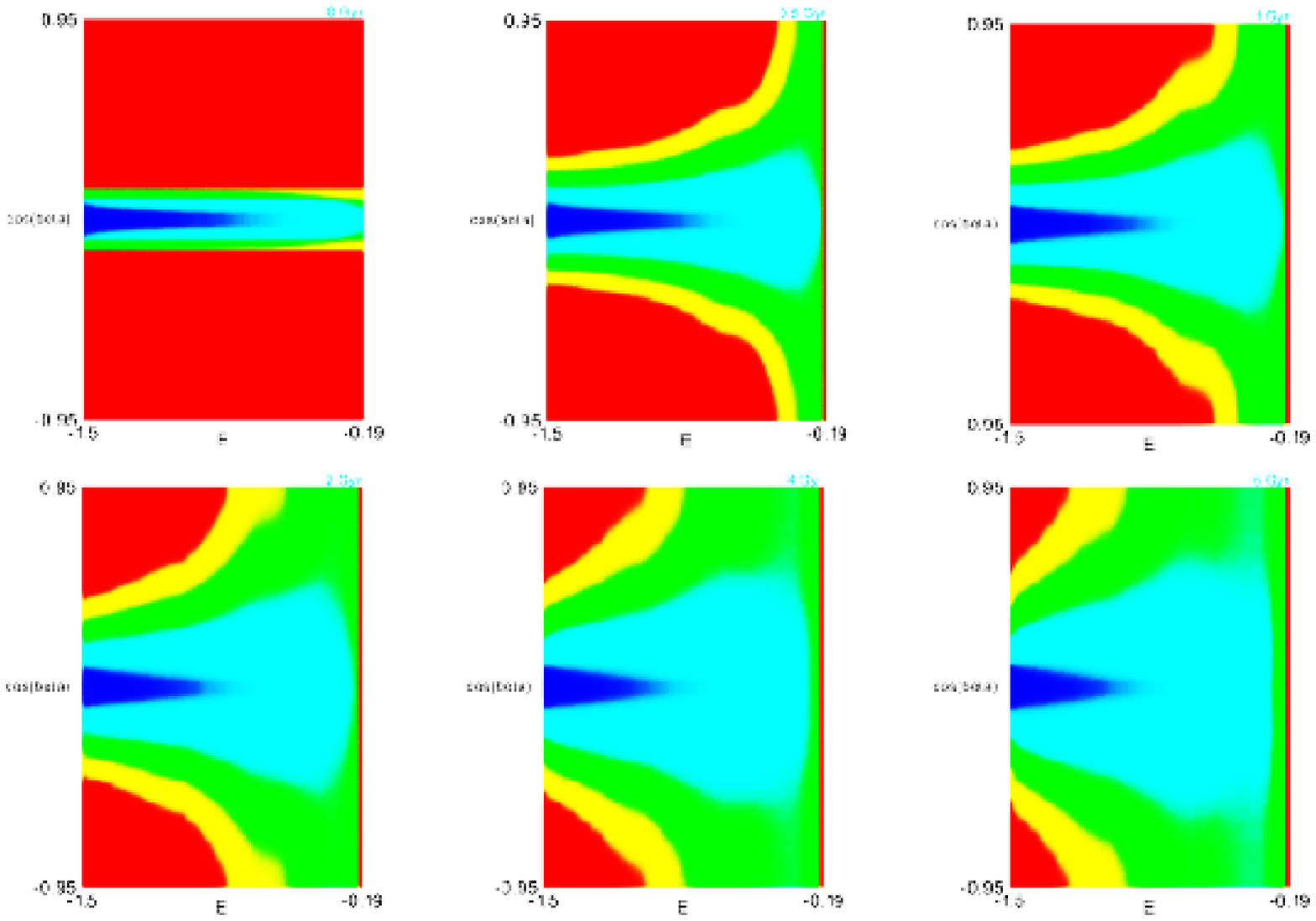}}
  \caption{As in Fig. \protect{\ref{fig:lmcdisk}} but showing the
    phase space density in the $E$ and $\cos(\beta)\equiv J_z/J$
    plane.  Orbits with higher binding energy are `heated' out of the
    disk plane with time.}
  \label{fig:lmcdisk_eb}
\end{figure}

The same dynamical couple that raises wakes in the halo affects the
LMC disk.  This changes the angular momentum of orbits at commensurate
frequencies and ``heats'' the disk.  To estimate the evolution, I
present a solution of the time-dependent collisionless Boltzmann
equation for orbits in a fixed potential.  This now linear PDE is
solved in a three-dimensional grid\footnote{E.g. $E$, $J$, $J_z$ or
  $E$, $J$, $\cos\beta\equiv J_z/J$ where $E$ is the orbital energy
  $J$ is the total angular momentum and $J_z$ is its $z$ component.}.
Figure \ref{fig:lmcdisk} shows both the cumulative mass distribution
above the disk plane and projected surface mass density.
Approximately 1\% of the disk mass has a height larger than 6 kpc and
10\% above 3 kpc after 1 Gyr.  This leads to a very thick disk or
flattened spheroid population.  These values are underestimates since
the self-consistent readjustment of the potential will lead to further
heating.  Although the projected profile appears unchanged for $T>1$
Gyr, this is an artifact of projection.  The orbits at low binding
energy are heated first and those at successively higher binding
energy as time goes on.  This is clearly seen in the
$E$--$\cos(\beta)$ projection phase space distribution (Fig.
\ref{fig:lmcdisk_eb}).  Also worthy of note is that $\log M(<Z)$ is
roughly linear with $Z$ which suggests an exponential profile.  A
similar profile has been reported for the RR Lyrae distribution in the
LMC halo (Kinman et al. 1991).  The sharp roll over at the tidal
radius is suggestive but may be an artifact of the particular
imposition of the tidal boundary.

\subsection{Microlensing}

An extended LMC halo\footnote{{\em Halo} here means any non-disk
  component.} can enhance the microlensing optical depth due to self
lensing.  We calculate the optical depth due to microlensing assuming
our spherical Gaussian halo model and the disk model from Wu (1994):
\begin{equation}
\rho_D = \rho_0 \: e^{-R/h} \sech^2 (z/z_0),
\end{equation}
where $h=1.6\;$kpc and $z_0=0.43\;$kpc. For now, the inclination of
the LMC disk (values range from $27^\circ$ to $45^\circ$) is ignored.
The mass of the LMC disk (out to $3^\circ$) of $4 \times
10^9\;M_\odot$ (De R\'ujula et al. 1995) implies $\rho_0 = 0.29\;
M_\odot/pc^3$. The halo mass, then, is $6 \times 10^9\;M_\odot$.  The
optical depth averaged along the line-of-sight is computed following
Kiraga \& Paczy\'nski (1994) with $\beta = -1$.

\subsubsection{Results}

First, assume there is no Galactic halo MACHOS: both source density
$\rho_s$ and deflector density $\rho_d$ include only the stellar halo
and the disk of the LMC. This gives the total optical depth due to LMC
self lensing of $1.7 \times 10^{-7}$; $55\%$ of this is due to halo
lenses.  A LMC halo mass of $1.2 \times 10^{10}\;M_\odot$ yields the
observed microlensing optical depth $2.9^{+1.4} _{-0.9} \times
10^{-7}$ (MACHO collaboration).  The prediction will still be
consistent with the reported value at the lower $1 \sigma$ limit if
the mass of the halo is $8 \times 10^9\;M_\odot$.  For a model
consisting only of LMC halo lenses, the optical depth is $1.9 \times
10^{-7}$ for halo mass $1.0 \times 10^{10} \;M_\odot$.  A halo mass of
$1.1 \times 10^{10}\;M_\odot$ yields a lensing prediction consistent
at $1\sigma$ level.

\begin{figure}[th]
  \mbox{\epsfxsize=2.5in\epsfbox{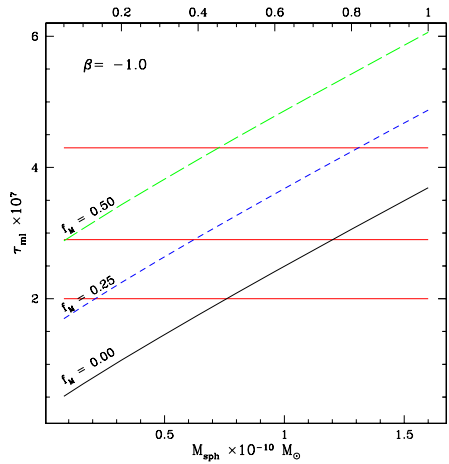}}
  \caption{Predicted microlensing depth $\tau$ vs. LMC spheroid mass.
    The three horizontal lines show the MACHO collaboration value and
    errorbars.  The curves show the predicted depth due to LMC lensing
    for Galactic MACHO fractions as labeled.}
  \label{fig:tau}
\end{figure}

If one now adds lensing by the Galactic MACHOs, the observed optical
depth is obtained for a MACHO fraction $f_{MACHO} = 0.25$ (assuming
the LMC mass of $1.0 \times 10^{10}\;M_\odot$).  This is marginally
consistent with the estimate $f_{MACHO} = 0.5^{+0.3} _{-0.2}$ obtained
by the MACHO collaboration.  These results are summarized in Figure
\ref{fig:tau}

The inclination of the disk can enhance the optical depth and
therefore can allow less massive halo to produce the observed optical
depth. This is similar to Gould's (1993) suggestion that the optical
depth can show asymmetric component for different lines-of-sight
through the LMC ($\sim 20\%$ asymmetry).

\section{Summary}

I hope to have convinced you that the natural dynamics of galaxian
halos provides a mechanism for carrying disturbances from the
extragalactic environment to the observable luminous disk. 

In particular, the orbiting LMC should have a profound effect on the
Milky Way.  For halo to disk ratios of ten to one and an estimate of
the LMC orbit based on its inferred space velocity, the halo wake can
excite a bending mode resulting in an observed disk warp.  A very
heavy or centrally concentrated halo can suppress the predicted warp
amplitude by detuning the near commensurability between the wake
pattern speed and bending mode.  In addition and independent of the
warp, the halo wake causes an in-plane $m=1$ distortion of roughly
20\% in the outer stellar disk.  This amplitude is similar to that
inferred from HI (Henderson et al. 1982).  I have most thoroughly
explored the LMC--Milky Way example presented here but fly-bys in
group and cluster environments cause similar excitation.

The importance of the LMC to Milky Way structure has led me to
consider the consequences to the LMC in more detail.  From the
LMC-centric viewpoint, the Milky Way is in orbit about the Cloud.
Because of the mass ratio, the Milky Way forcing is a huge effect on
the LMC!  The simplest effect is tidal limitation and leads to a mass
estimate $M_{LMC}\approx 2\times10^{10}\msun$.  In fact, there is
dynamical consistency between Milky Way halo rotation curve, LMC
rotation curve and LMC tidal radius which corroborates our large mass
estimate.  Furthermore, I have argued that the time-dependent forcing
perturbs the LMC well-inside of the tidal radius, exploiting the same
non-local resonant mechanisms that cause the Milky Way halo wake.
Because the LMC disk is inclined to its orbital radius, LMC disk
orbits may be torqued by Milky Way tide producing an extended and
rotating spheroid component.  This prediction is consistent with the
observation that nearly all LMC components have the disk kinematics
(Olszewski et al.  1996).  Finally, if the existence extended stellar
component is borne out, LMC--LMC gravitational microlensing rates will
be enhanced.  Conversely, if the LMC stellar component is confirmed to
be thin, we have misunderstood a basic feature of LMC dynamics.

\acknowledgments
This work was supported in part by NSF AST-9529328 and NASA/JPL 961055.

\end{document}